# ACO BASED ROUTING FOR MANETS

Mohammad Arif[1] and Tara Rani[2]

[1]Research Scholar, Department of Computer Engineering, Singhania University, Jhunjunu, Rajasthan, India
`arif_mohd2k@yahoo.com`
[2]Department of Computer Engineering, NICE College of Technology, Agra, Uttar Pradesh, India
`drtararani@rediffmail.com`

## ABSTRACT

*Mobile ad hoc network (MANET) is a collection of wireless mobile nodes. It dynamically forms a temporary network without using any pre existing network infrastructure or centralized administration i.e. with minimal prior planning. All nodes have routing capabilities and forward data packets to other nodes in multi-hop fashion. As the network is dynamic, the network topology continuously experiences alterations during deployment. The biggest challenge in MANETs is to find a path between communicating nodes. The considerations of the MANET environment and the nature of the mobile nodes create further complications which results in the need to develop special routing algorithms to meet these challenges. Swarm intelligence, a bio-inspired technique, which has proven to be very adaptable in other problem domains, has been applied to the MANET routing problem as it forms a good fit to the problem. In ant societies the activities of the individuals are not regulated by any explicit form of centralized control but are the result of self-organizing dynamics driven by local interactions and communications among a number of relatively simple individuals. This unique characteristic has made ant societies an attractive and inspiring model for building new algorithms and new multi-agent systems. In this paper, we have studied and reviewed Ant Colony based routing algorithms and its variants. Finally, a performance evaluation of the original ARA and the EARA is carried out with respect to each other.*

## 1. INTRODUCTION

Mobile ad hoc networks (MANETs) are networks that are made up of a set of mobile devices. There are no designated routers, meaning that all nodes can serve both as end points of data communication and as intermediate relay points or routers. Ad hoc networks must also support communication between nodes that are only indirectly connected by a series of wireless hops through other intermediate nodes. Mobile ad hoc networks (MANETs) are networks that are made up of a set of mobile devices. There are no designated routers, meaning that all nodes can serve both as end points of data communication and as intermediate relay points or routers.

A MANET routing algorithm should not only be capable of finding the shortest path between the source and destination, but it should also be adaptive, in terms of the changing state of the nodes, the changing load conditions of the network and the changing state of the environment. MANET routing algorithms can be classified into three categories as proactive, reactive or hybrid [7]. Proactive algorithms try to maintain up-to-date routes between all pairs of nodes in the network at all times. Examples of proactive algorithms are Destination-Sequence Distance-Vector routing (DSDV) and Optimized Link State Routing (OLSR) [12]. Reactive algorithms only maintain routing information that is strictly necessary: they set up routes on demand when a new communication session is started, or when a running communication session falls without route. Examples of reactive routing algorithms include Dynamic Source Routing (DSR) and Ad-hoc On-demand Distance-Vector routing (AODV) [21]. Finally, hybrid algorithms use both proactive and reactive elements, trying to combine the best of both worlds. An example is the Sharp Hybrid Adaptive Routing Protocol (SHARP) [9].The traditional routing protocols face many problems due to the dynamic behaviour and resource constraints in MANETs. To overcome this limitation, an approach to achieve such feature is to use a biologically-inspired mechanism.

The social organization of the ant is genetically evolved commitment of each individual to the survival of the colony. It is a key factor behind their success. Moreover, these insect societies exhibit the fascinating property that the activities of the individuals, as well as of the society as a whole, are not regulated by any explicit form of centralized control. The most successful and most popular research direction in ant algorithms is dedicated to their application to combinatorial optimization problems, and it goes under the name of Ant Colony Optimization metaheuristic (ACO).

ACO finds its roots in the experimental observation of a specific foraging behaviour of colonies of Argentine ants Linepithema humile which, under some appropriate conditions, are able to select the shortest path among the few alternative paths connecting their nest to a food reservoir. While moving, ants deposit a volatile chemical substance called pheromone and, according to some probabilistic rule, preferentially move in the directions locally marked by higher pheromone intensity.

Among the different works inspired by ant colonies, the Ant Colony Optimization metaheuristic (ACO) is probably the most successful and popular one. The ACO metaheuristic is a multi-agent framework for combinatorial optimization whose main components are: a set of ant-like agents, the use of memory and of stochastic decisions, and strategies of collective and distributed learning. It finds its roots in the experimental observation of a specific foraging behaviour of some ant colonies that, under appropriate conditions, are able to select the shortest path among few possible paths connecting their nest to a food site. ACO is based on the ant foraging behaviour, utilizing pheromone deposition as a means of evaluation for the travelled route. Rather than RREP and RREQ packets, 'forward' and 'backward ant' agents are sent across the routes, where the ant agents deposit pheromones at the nodes arrived. In the long term, this approach is used to determine the shortest path between the source and the destination.

In this paper we have focused on application of Ant Colony Optimization and various routing techniques for wired as well as for mobile ad hoc network. The Ant Algorithm mimics the behaviour of ants in nature while they are searching for food. The rest of the paper is organized as follows: Section 2 presents descriptions of routing and the ant system. Ant Colony Optimisation (ACO) and Ant Colony Routing (ACR) are discussed in Section 3 and Section 4 respectively. In Section 4 we have reviewed the analysis of performance of ARA and EARA. Lastly Section 6 summarizes our contributions.

## 2. ROUTING AND ANT SYSTEM

The core of any network control system is Routing which strongly affects the overall network performance. Routing deals with problem of defining path to forward incoming data traffic such that the overall network performance is maximized. At each node data is forwarded according to the content of routing table. In this sense, a routing system can be properly seen as a distributed decision system. A variety of different classes of specific routing can be defined according to the different characteristics of the processing, transmission components, traffic pattern and type of performance. The routing problem is composed of two parts: (i) the communication structure, which in a sense defines the constraints, and (ii) the traffic patterns that make use of this structure.

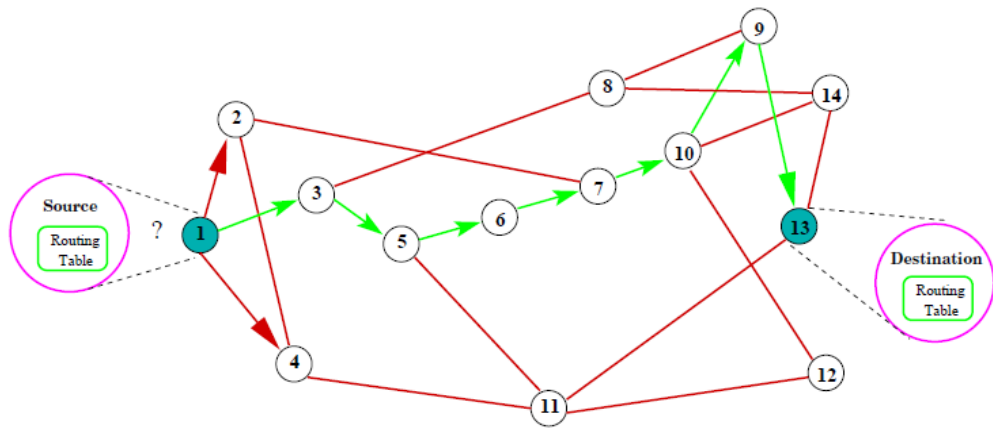

Figure 2.1. Routing in networks.

In figure 2.1, traffic data must be forwarded from the source node 1 to the target node 13. Several possible paths are possible. Each node will decide where to forward the data according to the contents of its routing table. One (long) path among the several possible ones is showed by the arrows.

In centralized routing algorithms, a main controller is responsible for the updating the routing tables and every routing decision for all the nodes. Centralized algorithms can be used only for small networks. In distributed routing systems, every node autonomously decides about local data forwarding. At each node a local routing table is maintained in order to implement the local routing policy.

Routing tables can be statically assigned or dynamically built and updated. In static routing systems, the path to forward traffic between pairs of nodes is determined without regard to the current network state. Once defined the paths to be used for each source-destination pair, data are always forwarded along these paths. In dynamic (or adaptive) routing system, the routing tables are dynamically updated according to the current traffic events and topological modifications (e.g., link/node failures, link/node addition/removal).

In shortest path routing the emphasis remains on minimum cost paths connecting all the node pairs in the network, where the paths are calculated independently for each pair. On the other hand, in optimal routing, the paths are calculated considering all the incoming traffic sessions.

## 2.1. Metrics for Performance Evaluation

The performance of a network is measured according to following metrics which depend on the types of services expected to be delivered by the network.

- Throughput: the number of correctly delivered data bits/sec. It is usually expressed as the sum of correctly delivered bits and/or packets over a specified time interval.

- End-to-end delay for data packets: the time necessary to a data packet to reach its destination node.

- Network resources utilization: considering both data and routing packets. Network resources commonly considered are the link capacities and the memory and processing time of the nodes. Network resources utilization is usually expressed as the used fraction of the overall available resources.

## 2.2. Ant in Nature

The main source of inspiration behind ACO and ACR is a behaviour that is displayed by certain species of ants in nature during foraging. It has been observed that ants are able to find the shortest path between their nest and a food source. The only way that this difficult task can be realized is through the cooperation between the individuals in the colony.

The key behind the colony level shortest path behaviour is the use of pheromone. This is a volatile chemical substance that is secreted by the ants in order to influence the behaviour of other ants and of it. Pheromone is not only used by ants to find shortest paths, but is in general an important tool that is used by many different species of ants.

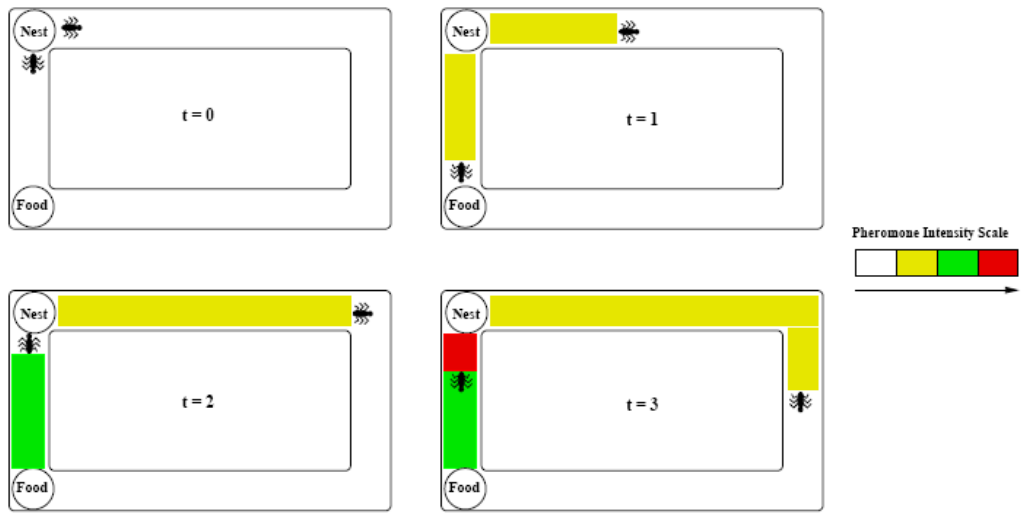

Figure 2.2: The shortest path mechanism used by ants. The different colours indicate increasing levels of pheromone intensity. From left to right and then from top to bottom, we see the situation in successive time steps [16].

Ants moving between their nest and a food source leave a trail of pheromone behind, and they also preferably go in the direction of high intensities of pheromone. We use the example situation depicted in figure 2.2 to explain how this simple behaviour leads to the discovery of shortest paths. In our example, there are two possible paths between the ant nest and the food source, one of which is considerably shorter than the other. The first ants leaving the nest have no information available. They therefore choose their movements randomly. This leads to approximately 50% of the ants choosing the short path and 50% choosing the long path. All moving ants leave a trail of pheromone behind. The ants going over the short path reach the destination earlier than those going over the long path. Moreover, they can return faster. This leads temporarily to a higher pheromone concentration on the shortest path. Subsequent ants leaving the nest are attracted by this higher intensity, and go therefore preferably also over the shortest path. As this process continues, the majority of the ants eventually concentrate on the shortest path. It needs to be pointed out however, that the behaviour of the ants is never deterministic, so that there will always remain a minority of ants that explore the longer path.

## 3. ANT COLONY OPTIMISATION

We have introduced a general framework for the design of routing algorithms (Ant Colony Routing), and reviewed the related work on ant-inspired routing algorithms. AntNet [18, 11, 20, 10, 14] and AntNet-FA [24] are two ACO algorithms for adaptive best-effort routing in wired datagram networks. On the other hand, Ant Colony Routing (ACR) is a general framework of reference for the design of autonomic routing systems [4]. ACR defines the generalities of a multi-agent society with the aim of providing a meta-architecture of reference for the design and implementation of fully adaptive and distributed routing systems for a wide range of network scenarios (e.g., wired and wireless, best-effort and QoS, static and mobile).

General characteristics of ACO algorithms for routing: The following set of core properties characterizes ACO instances for routing problems:

- provide traffic-adaptive and multipath routing,
- rely on both passive and active information monitoring and gathering,
- make use of stochastic components,
- do not allow local estimates to have global impact,
- set up paths in a less selfish way than in pure shortest path schemes favoring load balancing,
- show limited sensitivity to parameter settings

These are all characteristics that directly result from the application of the ACO's design guidelines, and in particular from the use of controlled random experiments (the ants) that are repeatedly generated in order to actively gather useful non-local information about the characteristics of the solution set (i.e., the set of paths connecting all pairs of source-destination nodes, in the routing case).

### 3.1. ACO Formulation

To formulate the ACO, the ants are modelled as artificial ants and the paths are represented as edges of a graph G. The formulation of ACO [13, 17, 15] as a combinatorial optimization problem can be done as follows:

$C = c_1, ..., c_n$ is a set of basic components. A subset S of components represents a solution of the problem; $F \subseteq 2^C$ is the subset of feasible solutions, thus a solution S is feasible if and only if S $\in$ F. A cost function f is defined over the solution domain, $f : 2^C \rightarrow R$, the objective being to find a minimum cost feasible solution S*, i.e., to find S*: $S^* \in F$ and $f(S^*) \leq f(S)$, $\forall S \in F$.

The search space S is defined as follows. A set of discrete variables, $X_i$, (i = 1, ..., n), with values $v_i^j \in D_i = \{v_i^1, ....., v_i^{|D_i|}\}$, is given. Elements of S are full assignments, i.e., assignments in which each variable $X_i$ has a value $v_{ji}$ assigned from its domain. The set of feasible solutions F is given by the elements of S that satisfy all the constraints.

In the Ant Colony Optimization, problems are usually modelled as a graph. Let G (V,E) be a connected graph with n = V nodes. Thus the components $c_{ij}$ are represented by either the edges or the vertices of the graph. The objective of the problem is to find a shortest path between the source node $V_s$ and destination $V_d$. Each edge of G maintains a value τ which denotes an artificial pheromone concentration value over that node which is modified whenever an ant transitions over it. To simulate the natural ant foraging process, three equations are used: Pheromone evaporation, Pheromone increase, and Path selection. If an ant currently at node i and transitions to node j:

$$\tau_{ij} = \tau_{ij} + \Delta\tau \qquad (3.1)$$

$\tau_{ij}$ is the artificial pheromone concentration over link j at i. The artificial Pheromones gradually evaporate over time, which is modelled by:

$$\tau_{ij} = (1 - \lambda) * \tau_{ij} \qquad (3.2)$$

Where $(1-\lambda)$ is called the pheromone decrease constant. At each node the ant has to make a decision about the next hop over which to travel. To simulate the exploratory behaviour of ants the artificial ant makes a stochastic decision based on probabilities of the next hop. The probability of an ant transitioning to node j from node i at node d, where Ni represents a set of neighbours, is calculated by the equation:

$$p^d_{ij} = \{ (\tau^k_{ij} / \sum_{j \in N_i} \tau^k_{ij}) \text{ if } j \in N_i \} \text{ and } \{d0, \text{ otherwise.}\} \qquad (3.3)$$

Where k is called the route selection exponent and determines the sensitivity of the ant algorithm to pheromone changes.

## 3.2. AntNet

AntNet is an ACO algorithm for distributed and traffic-adaptive multipath routing in wired best-effort IP networks. AntNet's design is based on ACO's general ideas as well as on the work of Schoonderwoerd et al. [1, 3], which was a first application of algorithms inspired by the foraging behaviour of ant colonies to routing tasks (in telephone networks). AntNet behaviour is based on the use of mobile agents.

Informally, the behaviour of AntNet can be summarized as follows:
- From each network node s mobile agents are launched towards specific destination nodes d at regular intervals and concurrently with the data traffic. The agent generation processes happen concurrently and without any form of synchronization among the nodes.
- Each forward ant is a random experiment aimed at collecting and gathering at the nodes non-local information about paths and traffic patterns. Forward ants simulate data packets moving hop-by-hop towards their destination. They make use of the same priority queues used by data packets.
- Ants, once generated, are fully autonomous agents. They act concurrently, independently and asynchronously. They communicate in an indirect, stigmergic way, through the information they locally read from and write to the nodes.
- The specific task of each forward ant is to search for a minimum delay path connecting its source and destination nodes.
- The forward ant migrates from a node to an adjacent one towards its destination. At each intermediate node, a stochastic decision policy is applied to select the next node to move to. The parameters of the local policy are: (i) the local pheromone variables, (ii) the status of the local link queues (playing the role of heuristic variables), and (iii) the information carried into the ant memory (to avoid cycles). The decision is the results of some tradeoff among all these components.
- While moving, the forward ant collects information about the traveling time and the node identifiers along the followed path.
- Once they have returned to their source node, the agent is removed from the network.

- Data packets are routed according to a stochastic decision policy based on the information contained in the data-routing tables. These tables are derived from the pheromone tables used to route the ants: only the best next hops are in practice retained in the data-routing tables. In this way data traffic is concurrently spread over the best available multiple paths, possibly obtaining an optimized utilization of network resources and load balancing.

### 3.2. AntNet-FA

In AntNet, forward ants make use of the same queues that are used by data packets. In this way they behave like data packets and experience the same travelling time that a data packet would experience. In this sense, forward ants faithfully simulate data packets. The problem with this approach is that, in case of congestion along the path that is being followed, it will take a significantly long time to the forward ant to reach its destination. Forward ants make use of high-priority queues as backward ants do, while backward ants update the routing tables at the visited nodes using local estimates of the ant travelling time, and not anymore the value of the time directly experienced by the forward ant. Forward ants make use of high-priority queues as backward ants do, while backward ants update the routing tables at the visited nodes using local estimates of the ant travelling time, and not anymore the value of the time directly experienced by the forward ant.

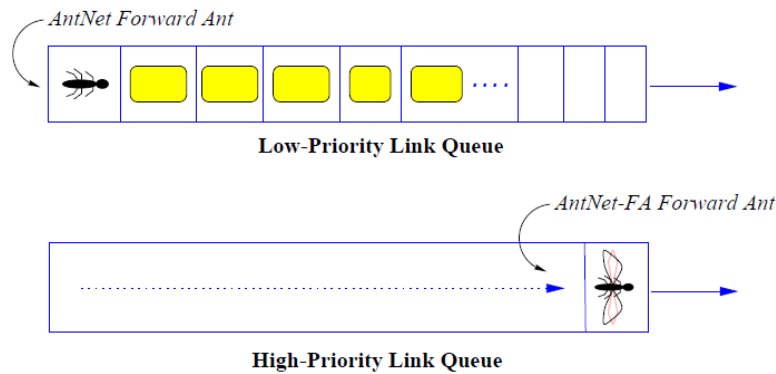

Figure 2: Forward ants in AntNet vs. forward (flying) ants in AntNet-FA

## 4. ANT COLONY ROUTING

The routing algorithms described so far have a flat organization and uniform structure: all the ants have the same characteristics and are at the same hierarchical level, while nodes are just seen as the repository of the data structures used by the ants.

ACR defines the routing system as a distributed society of both static and mobile agents. The static agents, called node managers, are connected to the nodes and are involved in a continual process of adaptive learning of pheromone tables that is, of arrays of variables holding statistical estimates of the goodness of the different control actions locally available. The control actions are expected to be issued on the basis of the application of stochastic decision policies relying on the local pheromone values. The mobile, ant-like agents play the role of either active perceptions or effectors for the static agents, and are generated proactively and on-demand. Node managers are expected to self-tune their internal parameters in order to adaptively regulate the generation and the characteristics of these subsidiary agents. In this way they are involved in two levels of learning activities. The active perceptions carry out exploratory tasks across the network and gather the collected non-local information back to the node managers. The effectors carry out ad hoc tasks, and base their actions on pre-compiled deterministic plans.

## 4.1. AntHocNet

In a MANET nodes can enter and leave the network at any time, as well as nodes can become unreachable becomes of mobility and limitations in the radio range. Therefore, in general it is not reasonable to keep at the nodes either a complete topological description of the network or a set of distances/costs to all the other nodes (even at the same hierarchical level) as it can be done, for instance, in the most of the cases of wired networks. These situations call for building and maintaining routing tables on the basis of reactive strategies possibly supported by proactive actions in order to continually refresh the routing information that might quickly become out-of- date because of the intrinsic dynamism of the network.

This is the strategy followed in AntHocNet [2, 8, 22, 23], which is a reactive-proactive multipath algorithm for routing in MANETs. The structure of AntHocNet is quite similar to that of AntNet-FA with the addition of some components specific to MANETs those results in the presence of several types of ant-like agents. In particular, the design of AntHocNet features: reactive agents to setup paths toward a previously unknown destination, per-session proactive gathering of information, agents for the explicit management of link failure situations (because of mobility and limited radio range the established radio link between two nodes can easily break). Node managers are not really learning agents, but rather finite state machines responding more or less reactively to external events. This is partially due to the fact that in such highly dynamic environments it might be of questionable utility to rely on approaches strongly based on detecting and learning environment's regularities. In the general case, some level of learning and proactivness is expected to be of some usefulness, but at the same time the core strategy should be a reactive one. This has been our design philosophy in this case.

## 4.2. ARA Algorithm

ARA is a purely reactive MANET routing algorithm. It does not use any HELLO packets [55] to explicitly find its neighbours. When a packet arrives at a node, the node checks it to see if routing information is available for destination d in its routing table. In ARA the route discovery is done either by the FANT (forward ant) flood technique [13] or FANT forward technique [12]. In the FANT flooding scheme, when a FANT arrives to any intermediate node, the FANT is flooded to all its neighbours. If found, it forwards the packet over that node, if not, it broadcasts a forward ant (FANT) to find a path to the destination. By introducing a maximum hop count on the FANT, flooding can be reduced. In the FANT forwarding scheme, when a FANT reaches an intermediate node, the node checks its routing table to see whether it has a route to the destination over any of its neighbours. If such a neighbour is found, the FANT is forwarded to only that neighbour; else, it is flooded to all its neighbours as in the flood scheme. In ARA, a route is indicated by a positive pheromone value in the node's pheromone table over any of its neighbours to the FANT destination. When the ant reaches the destination it is sent back along the path it came, as a backward ant. All the ants that reach the destination are sent back along their path. Nodes modify their routing table information when a backward ant is seen according to number of hops the ant has taken. When a route is found the packet is forwarded over the next hop stochastically according to equation 2.3. The results for the route discovery mechanism reveal an interesting trend. The FANT forwarding technique does better in situation of high mobility, that is, in situations having a lower pause time. However in cases of lower mobility, the FANT flood technique does better in the metrics of packet delivery ratio, throughput, delay and jitter. One more thing is being observed that in lower mobility situations, the FANT flood technique causes a lot of overhead, and increases the time required to find a route to the destination.

## 4.3. EARA Algorithm

ARA and AODV are compared by the author in [6] and ARA is found better than AODV. Since ARA is a reactive protocol, that is why it is used in such situations where mobility of nodes are higher. In this section, we have proposed the modifications to the algorithm by which the potential of ARA will increase in high mobility scenarios. Pheromone updates play a critical role in the performance of the ant algorithm. In ARA algorithm, initial pheromone value is computed by number hops during the route discovery. This method may not be suitable when nodes are mobile. Pheromone equations are classified in different categories. Two of them are the Classic pheromone filter, where route quality is not taken into consideration, for example the original ARA pheromone equation and the Gamma pheromone filter, which takes time and route quality into consideration.

Taking path quality into consideration we develop a type of Gamma Pheromone filter for ARA to update pheromone values as Gamma Pheromone filters show a better performance over the classic pheromone filters [25]. The modified pheromone update equation sets the initial value of the pheromone as:

$$\tau^d_{ij} = 2 / (hops + t) \qquad (4.1)$$

The pheromone update is done as per Equations (3.1) and (3.2). Where $\tau^d_{ij}$ denotes the pheromone concentration over link (i, j) for a destination "d". "t" denotes the time interval between the sending of a forward ant and the receipt of the backward ant, and hops is the total number of hops made by the ant. The inclusion of time in the equation creates a pheromone gradient from source to the destination point depending on the time it takes for the backward ant to reach the node that forwarded it. In the case of only FANT hops being taken into consideration, many paths with a similar gradient are formed; however the time metric creates a marked difference in the path gradient and thus the packet would be randomly forwarded over the path with the greatest pheromone gradient. This metric is thus expected to produce better results than if only number of hops is considered.

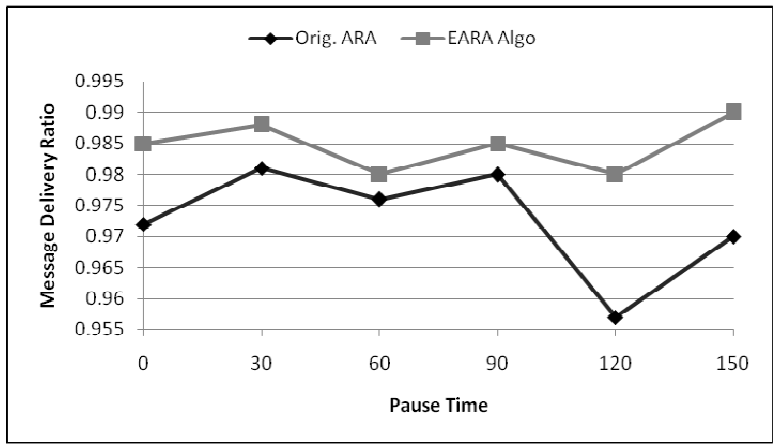

Figure 5.1: Message Delivery Ratio

## 4.4. Simulation Results

The performance is measured in terms of Packet Delivery Ratio, Throughput, End-to-End delay and Jitter for various values of pause time respectively. The observations show the efficiency of the modifications to the ARA algorithm. For the delivery ratio, throughput, and jitter metrics, the proposed ARA algorithm performs better than the original ARA algorithm. As shown in figure 5.1, 5.2, 5.3 and 5.4 all the performance metrics are enhancing except the delay.

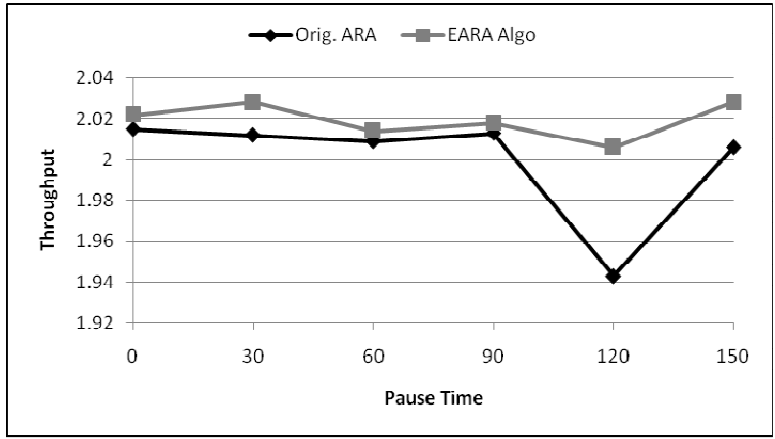

Figure 5.2: Throughput

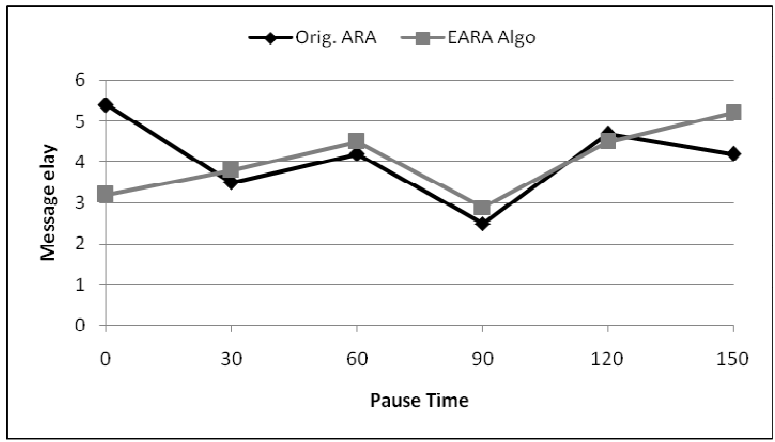

Figure 5.3: Message Delay

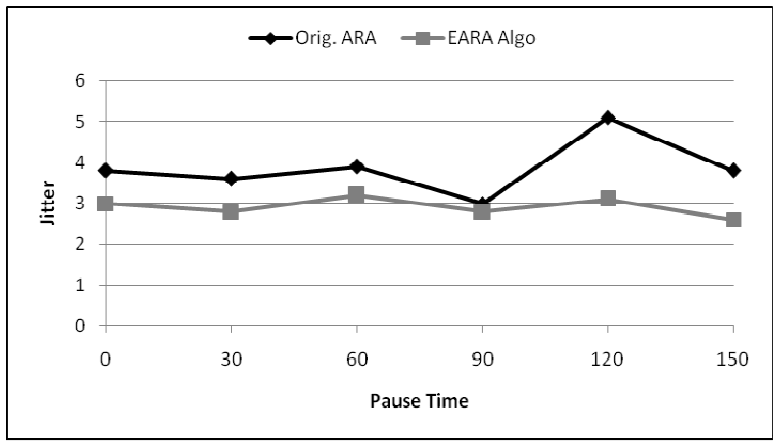

Figure 5.4: Jitter

## 5. CONCLUSION

The foraging behaviour of the ant colonies has been extensively investigated for more than 50 years and has been explored the remarkable trail systems achieved through robust, decentralized communication. Their collective intelligence has been shown to be one of the best examples of self-organization. Early research revealed their ability to detect shortest paths in static environments, whereas recent research discovered fundamental mechanisms in the foraging systems for the dynamic systems as well. In this paper we have surveyed the various techniques for wired and ad hoc networks. Through the simulations, some of the routing mechanism i.e. ARA and EARA is being analyzed and it has been found that EARA technique is working well in high mobility scenarios. The EARA is modified version of ARA algorithm and it is observed, through various simulation based experiments, that EARA performed better in comparison to the original ARA in terms of varying mobility. Ant Colony algorithms are very adaptive to the changing environments. However, their performance must be improved further and they must be able to solve the problems of heterogeneous networks.